\documentstyle[cite,aps,twocolumn,floats,psfig,prl]{revtex}

\newcommand{\psfigure}[1]{\centerline{\psfig{file=#1,width=8.5cm}}}

\begin{document}
%%%%%%%%%%%%%%%%
%\draft
\title{Formation of Chain-Folded Structures in Supercooled Polymer Melts}

\author{Hendrik Meyer$^{1,2}$
 and Florian M\"uller-Plathe$^2$ }
\address{$^1$Institut Charles Sadron, C.N.R.S., 6, rue Boussingault, 67083
  Strasbourg, France}
\address{$^2$Max-Planck-Institut f\"ur Polymerforschung,
        55021 Mainz, Germany}
% e-mail: hmeyer@ics.u-strasbg.de

\date{\today}
\maketitle

\begin{abstract}
  The formation of chain-folded structures from the melt is
  observed in molecular dynamics simulations resembling the
  lamellae of polymer crystals. Crystallization and subsequent
  melting temperatures are related linearly to the inverse lamellar
  thickness.  Analysis of the single chain conformations in the
  crystal shows that most chains reenter the same lamella by tight
  backfolds.  Simulations are performed with a mesoscopic
  bead-spring model including a specific angle bending potential.
  They demonstrate that chain stiffness alone, without an
  attractive inter-particle potential, is a sufficient  driving force
  for the formation of chain-folded lamellae.
\end{abstract}


%\paragraph*{Introduction}
For polymeric materials, the connectivity of the long-chain molecules
usually hinders the formation of perfect crystals.  Instead, most
crystallization conditions lead to partially crystalline structures
consisting of a stacking of chain-folded lamellae and amorphous regions.
There is nowadays wide agreement on the structure of this
semicrystalline phase\cite{Bar92}. However, the pathway which leads to
it and why it is taken is one of the unsettled questions in polymer
science. Actively debated ideas
\cite{Str2000epje,Lot2000epje,ChLiZh2000epje,Mut2000epje,RyFaTeOlPo99fd,OlPoMLTeRy98prl}
are intermediate mesomorphic phases\cite{Str2000epje} and the
possibility of a spinodal
mode\cite{ImKaKa93prl,RyFaTeOlPo99fd,OlPoMLTeRy98prl,MuWe2000pol} in
contrast to nucleation and growth.  Existing theories of polymer
crystallization \cite{HoMi97pol,SaGi86prl,ArGW92} are still considered
incomplete \cite{PoJa97pol,Mut2000epje,Str2000epje,Gei2000pol}.  Thus,
it is desirable to develop new tools for examining nucleation and
crystal growth, and especially its early stage.

Computer simulations generally have two advantages which may help to
understand the nonequilibrium process of polymer
crystallization.  First, the trajectories of all particles are
available. This can give information about single molecules which are
not accessible by experiments. Second, 
models of variable complexity can be simulated.
The use of successive levels of  approximations has the
advantage to give insight into the effect of certain interactions
separated from others.  In this spirit, this Letter explains how
chain-folded structures emerge from the supercooled melt when only the
excluded volume and the possible conformations of the polymer chains
are taken into account.

Many other simulation approaches have already been proposed in the
context of polymer crystallization.
Monte Carlo (MC) techniques were used to test theories of the lamellar
thickness with lattice models \cite{SaGi86prl,DoFr98prl,AnGW2000pol}.
\nocite{Doy2000pol} On a larger length scale, MC was successfully
applied to model 2D crystallization \cite{ReSo98prl}.
\nocite{SoRe2000jcp}
Molecular dynamics (MD) is often more realistic, but simulations have
been restricted either to few chains and short times
\cite{SuNoLiWu94aps}, or to structure formation of very short chains
\cite{Tak98jcp,FuSa98prl,ShYa2000jcp}, or to an isolated long chain
which collapses into a ``crystalline'' globule
\cite{LiJi99jcp,MuWe2000pol}.
% Specific boundary conditions were used
%to examine the folding of single chains \cite{Yam97jcp}.
%
Recent MD simulation studies which come closest to our work deal with
the crystallization of 20mers from the melt \cite{Tak98jcp}, of 10mers
in thin films \cite{ShYa2000jcp}, and of few long chains from solution
\cite{LiMu98jcp}.
Here, we present large scale MD simulations of structure formation
from the melt for a wide range of chain lengths and temperatures.  
We examine the homogeneous nucleation regime, since no walls or
starting nuclei are introduced.  In this Letter, we demonstrate in
particular that systems of chains long enough to fold can be
simulated  from the amorphous melt up to the
chain-folded lamella.  The individual particle trajectories 
are then used to analyze the nucleation process and the
chain folding in detail.

%\paragraph*{Computational Section}

We use a simplified version of a model for poly(vinyl alcohol) (PVA)
derived by a systematic coarse-graining procedure \cite{cgpaper2} from
fully atomistic simulations.  A coarse-grained bead represents one
monomeric unit of PVA. (In contrast, the often used united-atom models
\cite{LiMu98jcp,FuSa98prl,Tak98jcp} consider one bead per carbon.)  The
beads are connected by harmonic springs, and additionally interact by an
angle bending potential retaining information on the torsional states of
the atomistic backbone.
The nonbonded interactions are approximated by a Lennard-Jones 6-9
potential.  Furthermore, for the simulation of a dense melt, it is
sufficient to take only the repulsive part of the potential
\cite{KrGr90jcp}. This leads to some shift of energy and pressure
while it affects only little the qualitative behavior of the
system.  This approximation speeds up the simulation. But it also
switches off one possible driving force of crystallization, the
cohesive energy, which would increase with decreasing temperature.
Thus, when lowering  temperature, only the chain
stiffness will increase.

\begin{figure}
%\psfigurek{cg-mapping.eps}
%\psfigurek{ttc-angpotk.ps}
%\centerline{\psfig{file=cg-mapping-hoch.eps,width=2.2cm}
%  \psfig{file=ttb-angpotk.ps,width=6cm}}
\psfigure{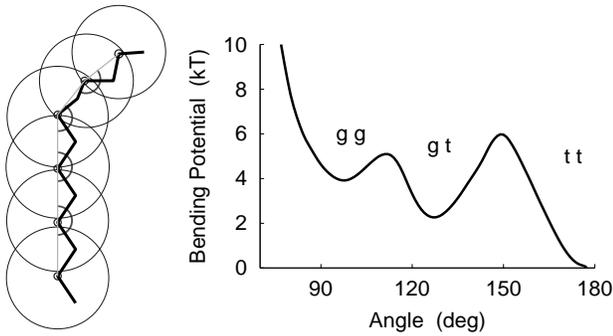}
\caption{Sketch of the coarse-grained model ``CG-PVA''.
  Each bead represents a monomer with two carbon atoms on the backbone.
  Successive beads are connected by harmonic springs.  Three
  successive beads interact via the shown angle bending potential.}
\label{f:angpot}
\end{figure}

Our model is defined by several parameters \cite{crystlong}.
Length scales are fixed by the mapping from atomistic simulations.
Units are reported in $\sigma=0.52$ nm corresponding roughly
to the chain diameter of PVA.
The bond length is  $0.5 \sigma =  0.26$ nm.
In these units, the density is about 2.1 monomers per $\sigma^3$.
The nonbonded potential between  monomers is
$V(r)/k_B =  -0.75/r^6 +0.53/r^9$
which is cut and shifted to zero at the minimum $r_{\rm cut} =1.02
\sigma$.  Nonbonded interactions between first and second
neighbors along each chain are excluded.  The angle potential  shown
in Fig.\ \ref{f:angpot}  has three minima at 180$^\circ$,
126$^\circ$, and 95$^\circ$  corresponding to trans--trans,
trans--gauche, and gauche--gauche conformations of the
atomistic backbone chain.
%
%%% FIG1
Note that this model, though initially fitted to represent PVA,
is also suitable for polyethylene which has almost the same
dimensions as PVA.
The equations of motion are integrated by the velocity-Verlet
algorithm using a timestep of 0.005 $\tau$ \cite{AlTi87}.  Temperature
is kept constant through a Langevin-thermostat with friction constant
$\Gamma = 0.5$.  Temperatures and energies are expressed in
dimensionless units with $m=k_B=1$; $T=1$ corresponds to the high
temperature phase of the amorphous melt.  Constant pressure is imposed
by the isotropic Berendsen manostat \cite{berendsen}.  The simulations
are performed in a cubic box with periodic boundary conditions.

\begin{figure}[tb]
\psfigure{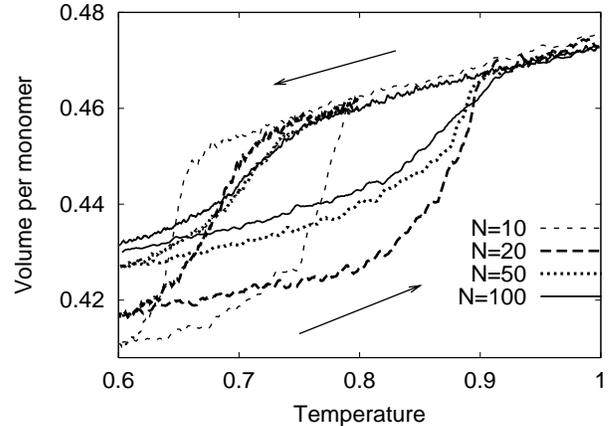}
\caption{Volume dependence in cooling/heating cycles for different
  chain lengths. The cooling rate is $5\times 10^{-6} \tau^{-1}$, the
  heating rate $2\times 10^{-5} \tau^{-1}$.}
\label{f:cycle}
\end{figure}

%\paragraph*{Results}
The model is studied for chain lengths $N=10$, 20, 50, and 100 by
two kinds of experiments. The first consists in cooling at constant
rate, the second is isothermal relaxation: the system is suddenly
quenched to a temperature $T_c$ below the melting temperature.  In
both cases we start with an amorphous configuration equilibrated at
$T=1$.  Figure \ref{f:cycle} shows the volume as a function of
temperature in cooling and heating cycles. It decreases with
temperature; a sudden drop of the volume is attributed to a phase
transformation.
%
%%%FIG2
Note the hysteresis.  Indeed, crystallization only takes place when a
stable nucleus is formed, i.e.\ at a certain
supercooling below the melting temperature.  In addition, the
hysteresis is enlarged due to the  fast cooling and heating
rates. Once formed, a crystal is also stabilized by the periodic
boundary conditions which lead to a higher melting temperature.
%
%%%FIG3
\begin{figure}[t]
%\psfigured{run100/plote-ang.ps}
%{run100/plot0-ang.ps}
\psfigure{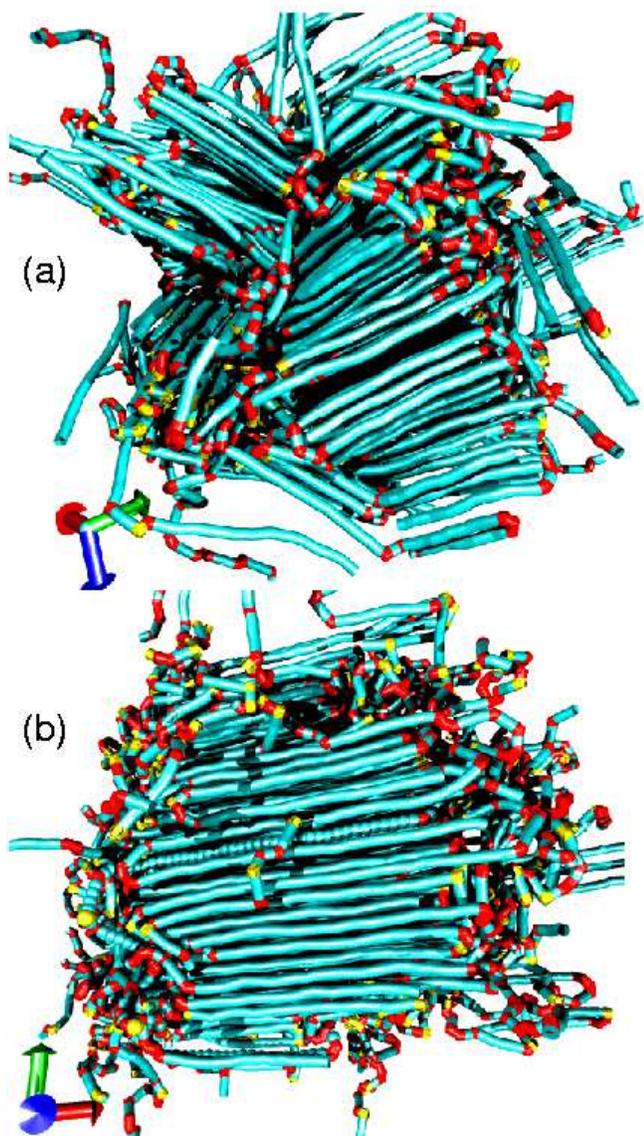}
\caption{(color) Two examples of structures obtained at 
  temperatures $T_c=0.7$ (a) and $T_c=0.8$ (b) with 192 chains of
  $N=100$ monomers.  Note the larger lamellar thickness at $T_c=0.8$.
  At $T_c=0.7$, two smaller domains with different orientation are
  formed. Color coding is according to the bond angle (cf. Fig.
  \protect\ref{f:angpot}): blue=tt, red=gt, yellow=gg.  The box length
  is about 10~nm. Note that periodic boundary conditions are applied;
  chains are not cut but mapped with their center of mass into the
  central box.}
\label{f:configs}
\end{figure}

Figure \ref{f:cycle} also shows that shorter chains begin to
crystallize at lower temperatures than longer chains. Similarly, the
decamers melt at lower $T_m$ than the icosamers. However, still longer
chains melt at about the same temperature as the icosamers.
Furthermore, the volume change for $N=50$ and $N=100$ is less
pronounced than for $N=20$. Both observations indicate that the
crystallinity of these samples is far from 100\%.
In fact, the shorter chains $N=10$ and $N=20$ always form extended
chain crystals. At temperatures below $T\approx 0.7$, crystallization sets in
immediately and small domains are formed similar to those reported in
Ref.~\cite{Tak98jcp}.  For smaller temperature jumps, the formation of a
single nucleus is observed which grows in regular layers.  More
details will be given elsewhere
\cite{crystlong}.

\begin{figure}[tb]
\psfigure{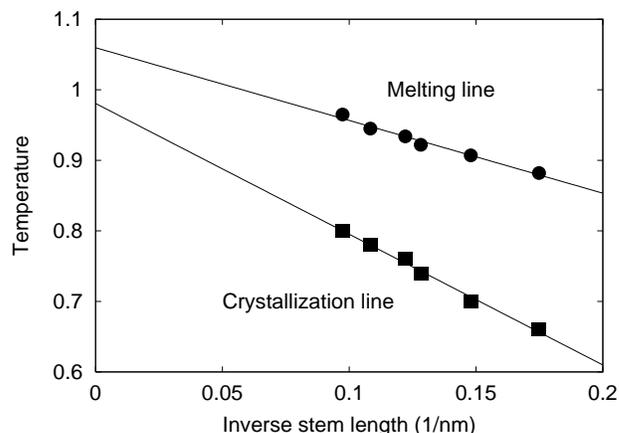}
\caption{Dependence of lamellar thickness on the crystallization temperature
  and melting temperature of the corresponding final structures. The
  lines are linear fits. Melting temperatures are rather overestimated
  due to finite size effects.}
\label{f:hw}
\end{figure}

Let us focus on the longest chains, $N=100$, in the following.
They always form  folded lamellar structures (see
Fig.~\ref{f:configs}).
The stems (the completely stretched parts)
are much longer at higher temperature.
Furthermore, the structure factor (not shown) indicates long range
order with Bragg peaks of the hexagonal lattice.  The structure factor
also has a peak at small wave vectors corresponding to the thickness
of the crystal lamella.
In Ref. \cite{crystlong} we discuss an alternative way of determining
the lamellar thickness with the bond orientation correlation function.
It gives an upper bound of the stem length.
Since most chains form tight backfolds (see below),
this  is a good estimate of the lamellar thickness.  Figure
\ref{f:hw} shows that the inverse of this average stem length
increases linearly with decreasing $T_c$, a relation observed
experimentally for most crystallizing polymers.  The same holds for
the melting temperature of the corresponding structures.
%
%%%FIG4
At $T_c=0.7$, most chains form 3 folds, at $T_c=0.74$ they have mostly
2 folds.  At $T_c=0.78$ the first nucleus grows with once-folded
chains; however, it continues to grow with a mixture of once
and twice folded chains which do not unfold during the simulation time.
At $T_c=0.8$, finally, most chains exhibit only one fold.

This can be summarized as follows: the higher the temperature the better
the crystal.  Though perhaps paradoxical at first sight, this is
generally observed for crystallizing systems and in particular for
polymers because the thermodynamically favorable extended chain crystal
is kinetically not accessible.  For a single chain, crystallizing with
many folds is much closer to the initial coil state in the melt than a
once folded or even extended chain. So, at lower temperatures, when the
mobility is lower and the nucleation probability is higher, the chains
will end up in crystals of short stems before having time to unfold.
The barrier between crystals of different fold lengths is high and is
overcome only with additional activation, for example, by annealing.

All simulation results reported so far are consistent with the
interpretation of well known experimental findings. This gives us
confidence in the model proposed.  On this basis, we may use the
simulation as a nanoscope and visualize the evolution of single chains
as well as the conformations of the chains in the crystal.
%
%%%FIG5
\begin{figure}[t]
\psfigure{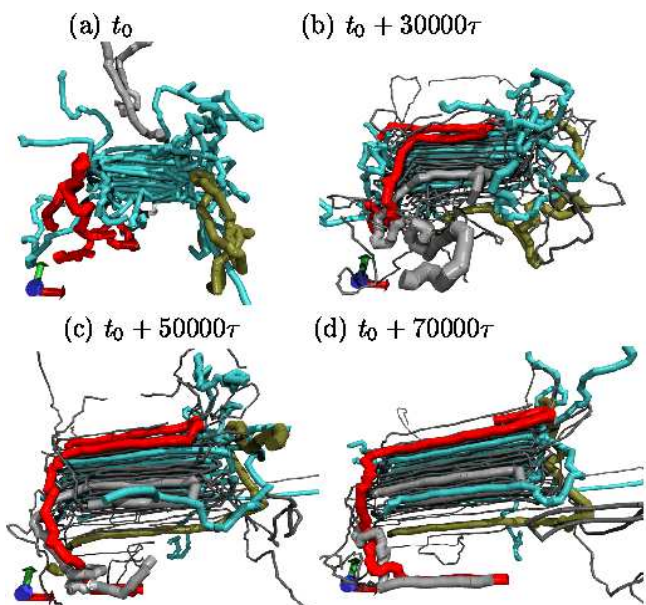}
%\centerline{(a) $t_0$ \hfil\hfil (b) $t_0+30000\tau$ \hfil}
%\psfigurekd{run100/plot0nn-0.pov.ps}{run100/plot0n-6.pov.ps}
%\centerline{(c) $t_0+50000 \tau$\hfil (d) $t_0+70000 \tau$ \hfil}
%\psfigurekd{run100/plot0n-10.pov.ps}{run100/plot0n-14.pov.ps}
\caption{(color) Snapshots of the crystallization at $T_c=0.8$.
  (a) The 13 chains participating in the initial nucleus are
  shown (blue).  The following snapshots contain 20 more chains (thin lines)
  which align to the crystal in the first 40000$\tau$.  (b) The
  red chain is partly attaching to the crystal front.  (c) The
  hairpin of this chain is growing to fill the whole lamellar
  thickness.  (d) Finally, the hairpin is vanishing: the chain glides
  until having only two stems in the crystal and a large bridge on the
  fold surface.}
\label{f:snap}
\end{figure}
Figure \ref{f:snap} illustrates snapshots from the crystallization at
$T_c=0.8$.  In this case a slightly different temperature protocol was
applied: the start nucleus was obtained by quenching the melt from
$T=1.0$ to $T=0.7$ over 10000 $\tau$. Then, the temperature was raised
to $T=0.8$ which ultimately leads to the structure in Fig.\ 
\ref{f:configs}b after 140000 $\tau$.  Initially, the nucleus grows
primarily in the direction of the stems (a,b). Later on, further stems
reorient and ``attach'' to the primary nucleus.  This ``attachment''
can start with arbitrary parts of the chain.  However, the chains move
until aligning their ends with the existing crystal.  In the sequence
of Fig.\ \ref{f:snap} one chain is highlighted which joins the crystal
in a complicated way. It attaches first via a hairpin which advances
until aligning with the total stem length of the crystal (c). However,
the other end of this chain is quite far. The chain slides along its
contour to finally annihilate the hairpin.

A quantitative analysis of the single chain conformations results in
3/4 of the folds in the final structure at $T_c=0.8$ being backfolds.
One third of them are hairpin folds, only 1/4 of the backfolds reenter
the crystal farther than at the 3rd neighbor position.  At lower
$T_c$, the fraction of backfolds in our samples was found to be around
60\%.  Due to domain formation (Fig.\ \ref{f:configs}a), up to 25\%\ 
of the folds link stems participating in different domains.
The distribution of the re-entry distances of backfolds is almost the same
for all crystallization temperatures.

%\paragraph*{Conclusions}
In summary, this Letter highlights results of a versatile  model
suitable for studying the chain-folding process of polymer
crystallization.  The chosen level of coarse-graining
retains enough information on the local conformations while the
speed-up due to the reduction of details is sufficient to follow 
nucleation and growth from the melt up to chain-folded structures.
Starting from the chemistry of PVA, the model finally consists only of
excluded volume interaction and a specific angle bending potential
favoring stretched conformations and allowing precise fold states;
there is no explicit attraction between monomers.
%Our results show that the interplay of these two ingredients is
%sufficient to yield chain-folded lamellar structures without an
%explicit attraction between monomers.
%
We understand the crystallization in our model as driven by an
energetic and an entropic component. When lowering temperature, energy
leads through the angle potential (Fig.\ \ref{f:angpot}) to an
increased chain stiffness (the persistence length of an isolated, free
chain passes from 3.4 monomers at $T=1.0$ to 6 monomers at $T=0.7$).
Entropy favors the parallel ordering as in liquid
crystal\cite{noteLC}, because the gain in translational entropy of the
ordered structure at lower temperature is possibly much larger than
the loss in conformational entropy.  The temperature dependent
kinetics then determines the form of the nonequilibrium structures of
chain-folded lamellae.
Yet, the qualitative chain length dependence of crystallization and
melting temperatures (Fig.\ \ref{f:cycle}) as well as the lamellar
thickness behavior (Fig.\ \ref{f:hw}) match well experimental
findings.  This underlines that the orientational and conformational
ordering play a major role for polymer crystallization.  It further
suggests that crystallization might pass through an intermediate step
of liquid crystalline order before locking in into the final crystal
structure \cite{OlPoMLTeRy98prl}, at least in a region in front of the
growth front. The model could now be used to address questions of
recent crystallization experiments in confined geometries 
and, with larger systems, the open
question of a spinodal initiation of  crystallization.

%\acknowledgments
We thank K. Kremer, J. Baschnagel, B.  Lotz, and J.-U. Sommer for
fruitful discussions. HM acknowledges financial support by European
Associated Laboratories MPI-P/ICS.
%\cite{Keller57,Fischer57}.

%%%%%%%%%%%%%%%%%%%%%%%%%%%%%%%%%%%%%%%%%%%%%%%%%%%%%%
%\bibliographystyle{aip}
%\bibliographystyle{prsty}
%\bibliography{bib/crystpol,bib/cg,bib/dpc,bib/divpol}

\begin{thebibliography}{10}
\bibitem{noteLC}
Note that our model chain is still flexible and has no intrinsicly stiff
parts as liquid crystalline polymers.
\end{thebibliography}

%%%%%%%%%%%%%%%%%%%%%%%%%%%%%%%%%%%%%%%%%%%%%%%%%%%%%%

%%%%%%%%%%%%%%%%%%%%%%%%%%%%%%%%%%%%%%%%%%%%%%
\end{document}